# A SIMPLE FORWARD SELECTION PROCEDURE BASED ON FALSE DISCOVERY RATE CONTROL[1]

By Yoav Benjamini and Yulia Gavrilov

*Tel Aviv University*

We propose the use of a new false discovery rate (FDR) controlling procedure as a model selection penalized method, and compare its performance to that of other penalized methods over a wide range of realistic settings: nonorthogonal design matrices, moderate and large pool of explanatory variables, and both sparse and nonsparse models, in the sense that they may include a small and large fraction of the potential variables (and even all). The comparison is done by a comprehensive simulation study, using a quantitative framework for performance comparisons in the form of empirical minimaxity relative to a "random oracle": the oracle model selection performance on data dependent forward selected family of potential models. We show that FDR based procedures have good performance, and in particular the newly proposed method, emerges as having empirical minimax performance. Interestingly, using FDR level of 0.05 is a global best.

**1. Introduction.** The problem of variable selection has attracted the attention of both applied and theoretical statisticians for a long time. Consider the widely used linear model, $Y = \mu + \varepsilon = X\beta + \varepsilon$, where $Y$ is a response variable, $X = (x_1, \ldots, x_m)$ is an $n \times m$ matrix of potential explanatory variables, which may include higher order terms and interactions. $\beta = (\beta_1, \ldots, \beta_m)$ is a vector of unknown coefficients, where some (or even most) of the coefficients may equal zero, and $\varepsilon = (\varepsilon_1, \ldots, \varepsilon_m) \sim N(0, \sigma^2 I)$ is a random error. We want to select a subset from the collection of the above $m$ explanatory variables to be used in predicting linearly the response variable, so that the mean square prediction error is minimized.

The common solution to this problem is to choose the appropriate subset $S$ by minimizing a model selection criterion of the form: $RSS_k + \sigma^2 k\lambda$, where $RSS_k = \sum_{i \in S}(Y_j - \hat{\beta}_i x_{ij})^2$ is the residual sum of squares from a least squares

Received February 2008; revised June 2008.

[1]Supported by US–Israel Binational Science Foundation Grant 1999441 and a US National Institute of Health grant.

estimator for a model with $k = |S|$ parameters and $\lambda$ is the penalization parameter.

However, the penalized methods utilizing a constant $\lambda$ are known to be ineffective when the real model size can vary widely, which is the case when the potential number of explanatory variables is large [Donoho and Johnstone (1994), Shen and Ye (2002) and Foster and Stine (2004)]. One major line of research addresses the problem by the use of nonconstant per parameter penalty function. The simplest suggestion is the Donoho and Johnstone universal threshold where $\lambda$ depends only on the size of the pool of variables over which the selection takes place, in the form of $\lambda_m = 2\log(m)$. More refined procedures use $\lambda_{k,m}$, that depends both on $m$ and on the size $k$ of the model considered.

In this paper we introduce a penalized model selection method that is based on a new adaptive false discovery rate (FDR) controlling procedure for multiple testing. Abramovich and Benjamini (1995) have pointed at the connection between model selection and multiple hypotheses testing, as setting coefficient to 0 amount to dropping the variables from the prediction equation. As Abramovich et al. (2006) (hereafter ABDJ) show, the basic FDR controlling multiple testing procedure can be translated into a penalty function, where the penalty per parameter increases with $m$, and decreases with $k$. The method in Benjamini and Hochberg (1995) (hereafter BH) was shown theoretically to have good asymptotic properties for orthogonal explanatory variables in sparse models. Johnstone and Silverman (2005) pointed at its limitation when the models are not sparse. The selection method proposed here should overcome these limitations, as it is based on an adaptive testing procedure that potentially addresses both high and low proportions of true hypotheses among the tested ones. The essence of the proposed procedure is to use penalized forward selection with the penalty function $\sigma^2 k \lambda_{k,m}$, with $\lambda_{k,m} = \frac{1}{k}\sum_{i=1}^{k} z^2_{(q/2)i/(m+1-i(1-q))}$, and stop at the first local minimum of the penalized RSS.

It is important to observe that all penalized methods can easily be computed using standard software such as R (S), SAS, SPSS, etc.: translating the penalty $\lambda_{k,m}$ into $p$-to-enter, and repeating the forward selection (or backward elimination) at different $p$-to-enter, allows even the more complex penalized methods, including the one proposed here, to be available to mass users of standard commercial statistical software (see the explicit algorithm in Section 3 and the example in Section 4).

In the above methods the penalty makes use of the number of parameters in the considered model, put differently the $l_0$ norm of the vector of parameters. A different line of research addresses the problem by moving to $l_1$—the sum of absolute sizes of the coefficients—to penalize the RSS. Methods like the Lasso [Tibshirani (1996)], LARS and Forward Stagewise [Efron et al. (2004)] and the Dantzig Selector [Candes and Tao (2007)] that take this approach, result in estimators that are not the least squares ones. In most of the proposed methods along the same line, to be reviewed in Section 2.3, there exist a tuning parameter that takes the role that $\lambda$ does in the sense that it controls the size of the model. The choice of the tuning parameter is based on cross validation ideas, and therefore these methods are computationally more intensive.

We wish to compare the performance of the proposed model selection method based on the adaptive FDR testing method to other FDR based penalized methods, as well as to other penalized methods. The comparison is made over a wide range of realistic settings: (1) nonorthogonal design matrices; (2) moderate and large, but finite pool of explanatory variables; (3) both sparse and nonsparse models, in the sense that (3a) they may include a small and large fraction of the potential variables (and even all), and (3b) both constant size coefficients and sizes that decay fast to 0.

Unfortunately, when it comes to analyzing performance for finite problems with nonorthogonal predictors the analytical tools are limited. Even the available few nice results [Birgé and Massart (2001)] are not useful for such comparisons. As many others do, we turn to a simulation study. However, all previous studies known to us were quite narrow in the number and the range of configurations studied, and suffered from the lack of a coherent way to discuss their results across configurations. Ours is a more comprehensive study, building upon configurations already offered as well as some new ones, and includes a quantitative framework for performance comparisons in the form of empirical minimaxity relative to a "random oracle."

We first review the nonconstant penalized model selection methods (Section 2) emphasizing the newly proposed one (Section 3). In Section 4 we analyze the model selection example raised by Efron et al. (2004) in order to demonstrate the results of the above methods as well as the computationally intensive LARS and FSR. Sections 5 and 6 describe the simulation study and comparison methodology respectively. The results are reported and analyzed in Section 7.

**2. Background.** As mentioned before, most of the penalized variable selection procedures share the same form of penalty function and differ only by the value of penalization coefficient, $\lambda$. They can be divided into three groups: (1) constant $\lambda$: forward selection (or backward elimination) with fixed $p$-to-enter ($p$-to-drop) where $\lambda = z^2_{(p-\text{to-enter})/2}$; AIC [Akaike (1973)] and Cp [Mallows (1973)] where $\lambda = 2 \approx z^2_{0.16/2}$; (2) $\lambda_m$ that is the function of the size of the set of variables $m$ over which the model is searched, for example the universal threshold of Donoho and Jonhstone. Note that the latter can also be viewed as a multiple testing Bonferroni procedure at level $\alpha_m = 1/\sqrt{\pi \log(m)}$, that is, $\lambda_m \approx z^2_{(\alpha_m/2)(1/m)}$ (with $0.19 \leq \alpha_m \leq 0.37$, when $10 \leq m \leq 10{,}000$); (3) $\lambda_{k,m}$ that is a function of both $m$ and the size of the considered model $k$, as described below.

2.1. *The BH-based penalty.* The FDR criterion in simultaneous testing is defined as the expected proportion of true null hypotheses rejected out of the total number of null hypotheses rejected [Benjamini and Hochberg (1995), hereafter BH].

The FDR-controlling testing procedure in BH runs as follows. Associated with $H_{0i}$ its $p$-value $p_{(i)}$, and $p_{(1)} \leq \cdots \leq p_{(m)}$ are the ordered $p$-values. Let $k = \max\{i : p_{(i)} \leq iq/m\}$. If such a $k$ exists, reject the $k$ hypotheses associated with $p_{(1)}, \ldots, p_{(k)}$, otherwise reject none. This procedure controls the FDR at level $q$, in fact at a level lower than $q$ by a factor of $m_0/m$, where $m_0$ is the number of hypotheses for which the null is true. When applying the FDR procedure in the orthogonal model selection situation, sorting the $p$-values is equivalent to choosing the variable sequentially (irrespectively whether in a backward or forward manner) according to standardized coefficients, since $(\frac{\hat{\beta}_k}{SE(\hat{\beta}_k)})^2 = \frac{RSS_{k-1} - RSS_k}{\sigma^2}$. Comparing the two-sided $p$-value $p_k$ with $kq/m$ is equivalent to comparing the above standardized coefficient squared with $z^2_{(k/m)(q/2)}$. Now $p_k \leq kq/m$ iff $RSS_k + \sigma^2 \sum_{i=1}^{k} z^2_{(i/m)(q/2)} \leq RSS_{k-1} + \sigma^2 \sum_{i=1}^{k-1} z^2_{(i/m)(q/2)}$. Adhering to the BH procedure one starts from the smallest standardized coefficient and stops at the first (rightmost) local minimum; using the same procedure in a forward selection manner and stopping at first (leftmost) local minimum is equivalent to the step-down version of the BH that also controls the FDR [Sarkar (2002)].

In view of the above, one can present these FDR procedures as a penalized method with a variable penalty factor

$$\lambda_{k,m} = \frac{1}{k} \sum_{i=1}^{k} z^2_{(i/m)(q/2)}. \tag{1}$$

ABDJ have made the above connection and showed that, for an orthogonal design matrix, the global minimum of the least squares penalized by (1) is asymptotically minimax for $l^r$ loss, $0 < r \leq 2$, simultaneously throughout a range of sparsity classes. The FDR parameter $q$ plays an essential role in the asymptotic minimaxity theory in ABDJ. Even if we allow variable $q_m$ for problem of size $m$, but in such a way that $q_m \to q < 0.5$, when $m \to \infty$, then it is a sufficient condition for asymptotic minimaxity to hold. In contrast, if the limit $q > 0.5$, asymptotic minimaxity is prevented.

It was also shown by ABDJ that, in the above setting, the difference between the locations of the rightmost minimum and the leftmost local minimum (bracketing the global minimum) is uniformly small. Therefore the asymptotic minimaxity of the global minimum holds for the other two as well. Moreover, numerically these indices are often identical.

NOTE. If the estimators have other distribution, $z^2_{(i/m)(q/2)}$ should be replaced by the same quantile of the appropriate distribution.

2.2. *Other nonconstant penalties.* For large $m$ and $k = o(m)$ the BH penalty is approximately the same as the following penalties:

$$\lambda_{k,m} \approx \frac{1}{k} \sum_{i=1}^{k} 2 \log(2m/qi) \approx 2 \log(2m/qk) = 2 \log(m/k) + 2 \log(2/q).$$

Several independent groups of researchers have proposed model selection procedures with penalties per parameter of similar form that are function of both $m$ and $k$:

(1) Foster and Stine (2004) arrived at a penalty $\lambda_{k,m} = \frac{1}{k} \sum_{i=1}^{k} 2 \log(\frac{m}{i})$ from information theoretic considerations.
(2) Tibshirani and Knight (1999) propose model selection using a covariance inflation criterion that adjusts the training error by the average covariance of predictions and responses on permuted versions of dataset. In the case of orthogonal regression, their proposal takes the form of penalized RSS, with the penalty being $\lambda_{k,m} = \frac{1}{k} 2 \sum_{i=1}^{k} 2 \log(\frac{m}{i})$. One may consider this much-simpler penalized form even in the nonorthogonal setting for which the original method was developed.
(3) Birgé and Massart have studied penalized model selection specifically designed to include penalties of the form $\lambda_{k,m} = 2 \log(Cm/k)$.
(4) Finally, George and Foster (2000) adopt an empirical Bayes approach, drawing the components from a mixture prior $(1-w)\delta_0 + wN(0, C)$ and then estimating $w$ and $C$ from the data. They argue that the resulting estimator penalizes the addition of a $k$th variable by a quantity close to $2 \log((m+1-k)/k)$.

2.3. *$L_1$ methods.* During the last few years, the attention in high dimensional linear regression problems has shifted to $l_1$ penalized approaches that do not rely on least squares estimators. Some examples of such methods are: Lasso, LARS, Forward Stagewise [Efron et al. (2004)] and the Dantzig Selector [Candes and Tao (2007)].

In the Lasso, one minimizes the residuals sum of squares subject to $\sum |\beta_i| \leq t$, where $t$ is a tuning parameter that governs the size of the model, and should be estimated by cross validation (in fact choosing $t$ large enough results in the least squares coefficients in the full model). Forward Stagewise is an iterative technique that begins with zero coefficients for all variables and the coefficients are modified in small steps, each time in the direction of the variable whose correlation with the current residuals is maximal (unlike regular forward selection where we look for the maximal absolute correlation between the current residuals and the variable adjusted for those already in the model). As in Forward Stagewise, LARS starts with all coefficients equal to zero, and finds the predictor most correlated with the response. It takes the largest step possible in the direction of this predictor until some other predictor has as much correlation with the current residual. At this point LARS proceeds in a direction equiangular between the two predictors, until a third variable enters the "most correlated" set, and so on. As shown by Efron et al. (2004), all three algorithms can be viewed as greedy forward stepwise procedures whose forward progress is determined by a compromise among the currently most correlated covariates.

Candes and Tao (2007) have proposed the Dantzig selector, that minimizes $\sum |\beta_i|$ subject to $\sup |Xr| \leq \lambda\sigma$, where $r$ is the current residuals and $\lambda$ some positive constant that controls the size of the model. Not surprisingly, they recommend $\lambda_m = \sqrt{2 \log m}$, and for sparse models the establish optimal $l_2$ rate properties. Bickel, Ritov and Tsybakov (2008a, 2008b) showed that the Lasso and the Dantzig Selector are approximately equivalent in terms of the prediction loss. However, when $m \gg n$, the bounds of the $l_2$ loss for two methods have a different numerical constants with no uniform dominance of one over the other.

## 3. The multiple-stage FDR procedure based penalty.
In sparse problems where the number of nonzero coefficients is small relative to the searched pool, the factor of $m_0/m$ is close to 1 and not that important. In nonsparse problems, adaptive FDR controlling methods were offered that have better performance as testing procedures [Benjamini and Hochberg (2000), Storey, Taylor and Siegmund (2004), Benjamini, Krieger and Yekutieli (2006) and many others]. They should have potentially better performance as model selection procedures. Unfortunately, not all such procedures can be translated into penalty functions and thereby be used in a simple way as model selection methods.

The multiple-stage FDR procedure in Benjamini, Krieger and Yekutieli (2006) re-estimates $m_0$ at the $i$th stage by subtracting from $m$ an estimated current bound on the number of correct rejections $i(1-q)$. It leads to constants of the form $\alpha_i = iq/(m+1-i(1-q))$, and the multiple-stage step-down testing procedure runs as follows:

(1) Let $i = 1$.
(2) If $p_{(i)} \leq \alpha_i$ reject the hypothesis associated with $p_{(i)}$, update $i$ to $i+1$ and repeat (2).

(3) If $p_{(i)} > \alpha_i$ stop.

Finner, Dickhaus and Roters (2009) show the asymptotic optimality of the testing procedure (and of asymptotically equivalent ones), under the mixture model of Genovese and Wasserman (2002). The FDR controlling property of the multiple-stage procedure for a finite number of hypotheses using independent test statistics was proved in Gavrilov, Benjamini and Sarkar (2009).

Using the same idea as outlined in Section 2.1 for the orthogonal design, defining the penalty factor $\lambda_{k,m} = \frac{1}{k} \sum_{i=1}^{k} z^2_{(q/2)(i/m+1-i(1-q))}$ and searching for the first minimum on the forward path is identical to applying the original multiple-stage procedure. In the nonorthogonal case we use the same penalty. As we still have $(\frac{\hat{\beta}_k}{SE(\hat{\beta}_k)})^2 = \frac{RSS_{k-1} - RSS_k}{\sigma^2}$, we practically test the coefficients sequentially using the FDR correction, so the $k$th variable is added to the current model if its standardized coefficient given the previous $k-1$ variables already entered into the model is larger than $z^2_{\alpha_k/2}$.

It is important to emphasize that the above procedure is not the same as computing all marginal $p$-values and applying the multiple testing procedure, nor is it equivalent to fitting the model with all variables and testing using these $m$ $p$-values. Applying forward selection on dependent predictors does not guarantee that the $p$-values for testing the additional coefficient are entered in the same increasing order. Moreover, all $p$-values are updated at each step of the forward selection, in contrast to the original multiple-stage procedure where the sequence of $p$-values is constant and ordered.

Figure 1 presents the comparisons between the penalty factor of BH and multiple-stage procedures. Because of their similarity for sparse models, it can be expected that the multiple-stage procedure enjoys the good asymptotic properties of the BH procedure for sparse models. However, for non-sparse models the penalty of multiple-stage procedure is smaller, allowing richer models to be selected when this is indeed the case ($m_0/m \ll 1$).

The multiple-stage procedure in its penalized form can be computed using forward selection (by any statistical software) in the following way:

(1) Let $i = 1$.
(2) Run the forward selection with $p$-to-enter equals to $qi/(m+1-i(1-q))$.
(3) Let $i'$ be the size of the model selected in (2).
(4) If $i' = i$ stop; otherwise let $i$ be $i'$ and repeat (2).

In fact, any penalized method with $\lambda_{k,m}$ decreases in $k$ can be computed in a similar way to find first/last local minimum.

NOTE 1.   In the multiple-stage procedure, the hypotheses can be rejected even if its $p$-value is greater than 0.5. This is not surprising, since all the hypotheses are tested simultaneously and the control of FDR allows a few erroneous rejections if many correct rejections have already been made. If one does not want to reject hypotheses with large $p$-values, a constraint may be added that the hypotheses cannot be rejected unless $p_i \leq C$, leading to $\lambda_{k,m} = \frac{1}{k} \sum_{i=1}^{k} z^2_{\min((q/2)i/(m+1-i(1-q)),C)}$.

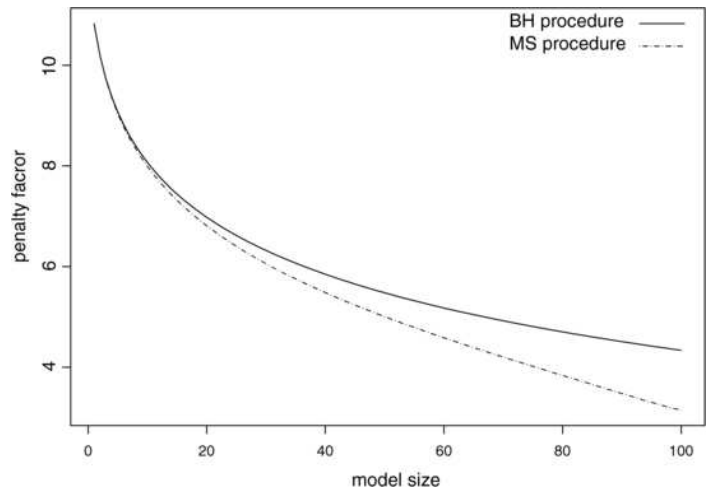

Fig. 1.   *The penalty factor of BH and multiple-stage procedures at FDR level* 0.05.

NOTE 2.   The implementation of the procedure in R (and Splus) is available at http://www.math.tau.ac.il/~ybenja.

**4. Modeling the diabetes data [Efron et al. (2004)].**   We now want to demonstrate our model selection procedure, as well as other penalized model selection procedures on the data first used in Efron et al. (2004), to illustrate their model selection method, LARS, and then by Wu, Boos and Stefanski (2007) to illustrate their false selection rate (FSR) method that is based on adding pseudo variables to the real dataset. Ten baseline variables, age, sex, body mass index, average blood pressure and six blood serum measurements, were obtained for each of $n = 442$ diabetes patients, as well as the response of interest, a quantitative measure of disease progression one year after baseline. The statisticians were asked to identify the important baseline factors in disease progression and to construct the model that produces accurate baseline predictions of response for future patients.

The data were standardized so that the baseline variables have mean 0 and unit length, and that the response has mean 0. Then, two sequences of nested models were produced by forward selection, one using only 10 main effects, and another using 10 main effects, 45 two-way interactions and 9 squares (each baseline variable except the dichotomous variable SEX). We do not force the hierarchy principle, and thus interaction effects can enter before main effects do.

The selection procedures were applied to the data, once searching for main effects only and once searching for main effects and interactions. In both cases we use the penalty of FDR procedures with $q = 0.05$. (We get the same models also if we use $q = 0.10$.) Standard errors were estimated from the full model.

In the main effect forward selection, the variables enter the model in the following order: BMI, S5, BP, S1, SEX, S2, S4, S6, S3, AGE. Both FDR procedures, AIC, DJ and forward selection with 0.05 select into the model the first 6 variables, exactly like FSR method. TK method stops after selecting 8 variables, BM—3 variables, FS—all 10 variables. The subset chosen by LARS is quite close to that of chosen by FDR methods: instead of S2, S3 and S6 are included, so this subset contains 7 variables.

In the quadratic model, the FDR procedure, DJ and TK select for the model 7 variables—5 main effects and 2 interactions. AIC method stops after 16 variables, forward selection with 0.05 and FS—13 variables and BM—2 variables. (For comparisons of all penalized procedures with LARS and FSR, see Table 2 in Supplementary Materials [Benjamini and Gavrilov (2009)].)

It is interesting to note that in both cases our results are identical to those of Wu, Boos and Stefanski (2007), but in the FDR based selection procedures there is no need to estimate parameters via bootstrapping, and therefore, they are faster and easier to implement. In fact, as noted in Introduction, any user of standard statistical software can use them.

As Wu, Boos and Stefanski (2007) noted, for the quadratic model the seven variables selected by the FSR, and as shown here by both FDR procedures, is the subset of the 16 variables included in the LARS model. The LARS model has $R^2 = 0.549$ while the others have $R^2 = 0.534$; the increase of 0.0153 comes at the expense of having nine additional variables in the model, indicating that LARS over-fits in this case. This tendency of LARS (and the Lasso) to overfit the model was noted by Bickel, Ritov and Tsybakov (2008a, 2008b).

Since the true model is not known in this case, we may choose to compare the performance of multiple-stage procedure and LARS, based on fivefold cross validation. This was further repeated 100 times, each time using a different (random) partitioning to the five subsets. The difference between the performances of the methods in terms of mean square prediction error was very small and not significant ($p = 0.98$). For the main-effects model, the FSR and FDR procedures produce very similar models to that of LARS, with nearly identical $R^2$. In summary, this example demonstrates that penalized selection methods works at least as well as the more complex ones.

Finally, we use the example to demonstrate how the MS selection procedure of a quadratic model, as described in Section 3, can be performed using standard software. We start by running the linear forward regression in SPSS with $p$-of-entry $= 0.00078$ that is calculated from $\alpha_1 = q/(m+q)$, where $m = 64$ and $q = 0.05$. This search stops after 5 variables enter the model (including the intercept). Now we update the $p$-of-entry to $\alpha_5 = 5q/(m+1-5(1-q)) = 0.004$. The new search stops with 8 variables in the model. Updating the $p$-of-entry to $\alpha_8 = 8q/(m+1-8(1-q)) = 0.007$, we again get the model with 8 variables (including the intercept). Since the model size no longer changes, the search is stopped.

## 5. Simulation study.

A simulation study was performed to compare the efficiency of the FDR controlling procedures and some other penalty based selection procedures. Table 1 in Supplementary Materials [Benjamini and Gavrilov (2009)] describes the selection procedures and configurations studied by various authors. The scope of each simulation by itself, appearing in the literature, is quite limited. Even when considering them jointly, they do not span a wide configuration space. The number of potential explanatory variables is at most 50. The most widely studied structure of nonzero coefficients is "all at a constant size chosen to yield a fixed correlation." In most studies the proposed method is compared with AIC and BIC, but with none of the other nonconstant adaptive penalties.

### 5.1. *The procedures.*

Ten model selection procedures were compared with the random oracle performance: (1) The Linear Step-up procedure in BH—FDR; (2) the two-stage FDR procedure (the modified version) in Benjamini, Krieger and Yekutieli (2006)—TSFDR; (3) the multiple-stage procedure in Benjamini, Krieger and Yekutieli (2006) and Gavrilov, Benjamini and Sarkar (2009)—MSFDR; (4) forward selection procedure—FWD; (5) the procedure in Foster and Stine (2004)—FS; (6) the procedure in Tibshirani and Knight (1999)—TK; (7) the procedure in Birgé and Massart (2001)—BM; (8) the procedure in George and Foster (2000)—GF; (9) the universal threshold in Donoho and Johnstone (1994)—DJ; (10) AIC (Cp) procedure. The first four procedures were implemented at levels $0.05, 0.10, 0.25$ and $0.50$, resulting in 22 model selection procedures investigated.

### 5.2. *The configurations investigated.*

Simulations are conducted with correlated and independent explanatory variables. A random sample $\{(Y_i, X_i)\}_{i=1}^n$ is generated according to $Y_i = X_i\beta + \varepsilon_i, \varepsilon_i \sim N(0, \sigma^2)$. The number of potential explanatory variables $m = 20$, 40, 80 and 160, the number of real explanatory variables (with nonzero coefficients) $p = \sqrt{m}, m/4, m/3, m/2, 3m/4$ and $m$. We denote these in the figures by $p =$ "1" to "6," from the sparsest model to the full one. The number of observations $n$ is always $2m$, and $\sigma = 1$. For each $m$, three design matrices were generated for three values of $\rho = 0.5, 0, -0.5$, from $N(0, \Sigma_{m \times m})$ with $\Sigma_{m \times m}$ whose diagonal elements are 1 and whose $ij$th element is $\rho^{|i-j|}$ for $i \neq j$. For each configuration of the parameters defined above, three different choices of $\beta$ are considered:

(1) $\beta = c(m)\{1/\sqrt{i}\}_{i=1}^{p}$;

(2) $\beta = c(m)\{p/(mi)\}_{i=1}^{p}$ for $p = m, 3m/4, m/2, m/3, m/4$, so the smallest nonzero $\beta$ will be always $c(m)/m$. For $p = \sqrt{m}$ $\beta$ was chosen uniformly on the interval $c(m)(1/m, 1)$;

(3) $\beta = \{c(p)\}_{i=1}^{p}$, where the value of $c(p)$ is chosen to give a theoretical $R^2 = \frac{\beta' X' X \beta}{\beta' X' X \beta + n} = 0.75$. The value of $c(p)$ decreases as $p$ increases.

We denote these three types of coefficients in the tables and figures as "1," "2" and "3."

Configuration (2) differs from configuration (1) both in the rate of decrease and in the fact that the minimal $\beta$ is constant across $p$. In both configurations $c(m)$ is chosen so that the standard error of the least squares estimators of the coefficients in the true model will be approximately equal for all values of $m$. Configuration (3) was chosen as it had been used by several authors as a test case in their simulation studies [George and Foster (2000); Shen and Ye (2002)]. The intercept for all configurations was chosen $\beta_0 = 10$, so that it is practically always included in the model.

In our initial simulations (not reported here), two additional configurations of coefficients and different $c(m)$ factors were included. The configurations of coefficients presented here expose extreme performances for all compared selection procedures for better or worse, and thus suffice for minimax evaluation.

5.3. *Computational procedures.* The performance of each model selection procedure in each configuration was measured by its mean square predicted error averaged over 1000 replications. The computational task involved in the current study was beyond achievement using a single computer within a reasonable time frame. We therefore utilized the software tool for computer-intensive jobs, Condor (http://www.cs.wisc.edu/condor/), which allows running serially and in parallel jobs, using the large collections of distributively owned computing resources. In our case we used the approximately 80 computers in our Students Computer Lab at the Schools of Computer Sciences and Mathematical Sciences at Tel Aviv University (operating on LINUX).

**6. Comparison methodology.** In problems of even medium size it is technically impossible to apply an exhaustive search over all possible subsets, therefore we limit the search for the best model over a selected path of nested models. We adhere in this work to the path generated by forward selection, as the motivating testing method starts with the most significant coefficient, and then adds one-by-one more coefficients. (We also have in mind current large applications where the number of variable considered may be larger than the number of observations available, and the use of a forward path is more natural.) Admittedly, such a path is data dependent, and need not contain the best overall model. If the produced path is far from being optimal, the performance of all tested methods on this path is beforehand restricted, since all penalized methods propose only the stopping rule with no corrections for "bad" path. By restricting attention to the same path we manage to separate the task of producing the good path of nested models from the task of stopping at the best model on this path.

Since in practice the configuration of parameters is unknown we would like to summarize performance across all configurations studied. For each configuration of parameters the performance of of a model selection method was measured by its MSPE: $MSPE_k = E[(\beta X_i - \hat{\beta}_k X_i)'(\beta X_i - \hat{\beta}_k X_i)] = \sigma^2 k + (\beta_2 X_2)'(I - X_1(X_1'X_1)^{-1}X_1')(\beta_2 X_2)$ where $X = [X_1, X_2]$, $X_1$ are the variables in the current model, $X_2$ are the variables out of the current model and $\beta_2$ is the vector of true coefficients corresponding to $X_2$. However, the comparison between estimators cannot be done directly in terms of MSPE, as MSPE values in one configuration can be vastly different from MSPE values in the other, so we have to work in terms of relative performance at the configuration level. Finally, one can revert to estimator whose maximum relative MSPE is minimal over the configurations studied, that is, the method that has the empirical minimax performance.

The benchmark for relative error can be set at the best estimator at that configuration among the tested ones, as was done in robustness studies [Andrews et al. (1972)]. This leads to results that are strongly dependent on the initial pool of estimators studied, so we avoid this option.

Using the ideas similar to the "oracle" of Donoho and Johnstone (1994), we define the "random oracle" as a benchmark. The random oracle calculates the above theoretical MSPE along the random path produced by forward selection, making use of the true values of the coefficients, and chooses the model with minimal MSPE. Then, we compare the performance of each procedure on the same random sequence relative to the performance of the random oracle. This comparison attenuates the stochastic component in the comparison that depends on the chosen path of nested models and gives us a common basis for all procedures even if the chosen path is far from including the global optimum.

The errors for the relative performance of each method to the random oracle were approximated from the standard errors of MSPE from the simulation results, using estimators for standard error of ratios [Cochran (1977)].

**7. The simulation results.** Figure 2 presents the configuration where $m = 80$, $\rho = 0$. For $\beta$-type = "1" $(\frac{1}{\sqrt{i}})$ and $p =$ "1" $(p = \sqrt{m})$ using $q = 0.05$ is best for all FDR procedures, while under the same setting for $p =$ "6" $(p = m)$, $q = 0.25$ is the best. Figure 3 (as well as Tables 3–6 in Supplementary Materials [Benjamini and Gavrilov (2009)] presents the MSPE values of three studied FDR procedures (BH, TS and MS) at level 0.05 relative to those of the random oracle estimator over all configurations for each $m$ and $\rho$ separately. Based on our simulation results, and on theoretical results of ABDJ, $q = 0.50$ never leads to the best performances. The FDR control parameter $q$ plays a role in achieving the minimal loss in each specific configuration studied. As expected, the small values of $q$ perform better in "small" models and vice versa. However, the information available for the model builder is only on $m$, sometimes on $\rho$, rarely vaguely on $p$ and never on the structure of the coefficients. Therefore, if we choose $q$ using a minimax consideration over the last two, we get Table 1 which summarizes the preferable values of $q$ for the FDR procedures.

It is evident that smaller values of $q$ (0.05 and 0.1) perform better than $q = 0.25$ in the studied range. In addition, while the optimal choice of $q$ for BH procedure varies across different $m$ values, within the same size $m$ it varied only in the configuration $\rho = -0.5$. For the proposed MSFDR the optimal values of $q$ are more stable as $m$ changes.

As it follows from Figure 2, the BH procedure at 0.05 has an advantage over the two adaptive FDR procedures at 0.05 only for "small problems," in particular for $m = 20$ $\beta$-type = "1" $(1/\sqrt{i})$ and $p$ is up to $m/2$ and $\beta$-

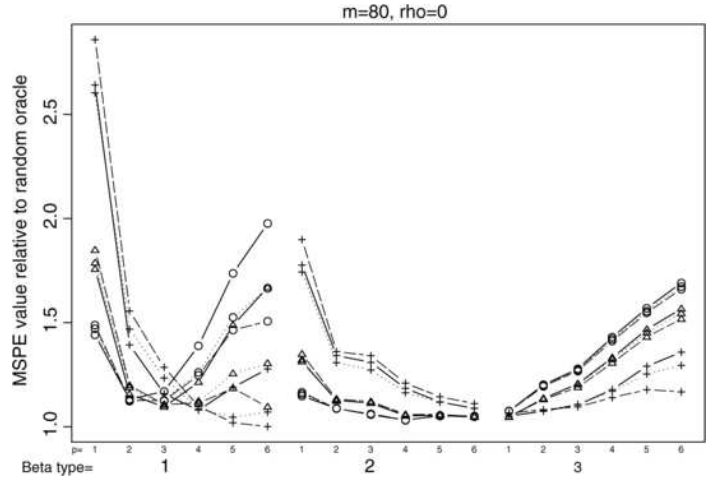

Fig. 2. *The MSPE values relative to the random oracle of FDR procedures: BH—solid line; TSFDR—dotted line; MSFDR—dashed line at 5% ($\circ$), 10% ($\triangle$) and 25% (+) FDR level.*

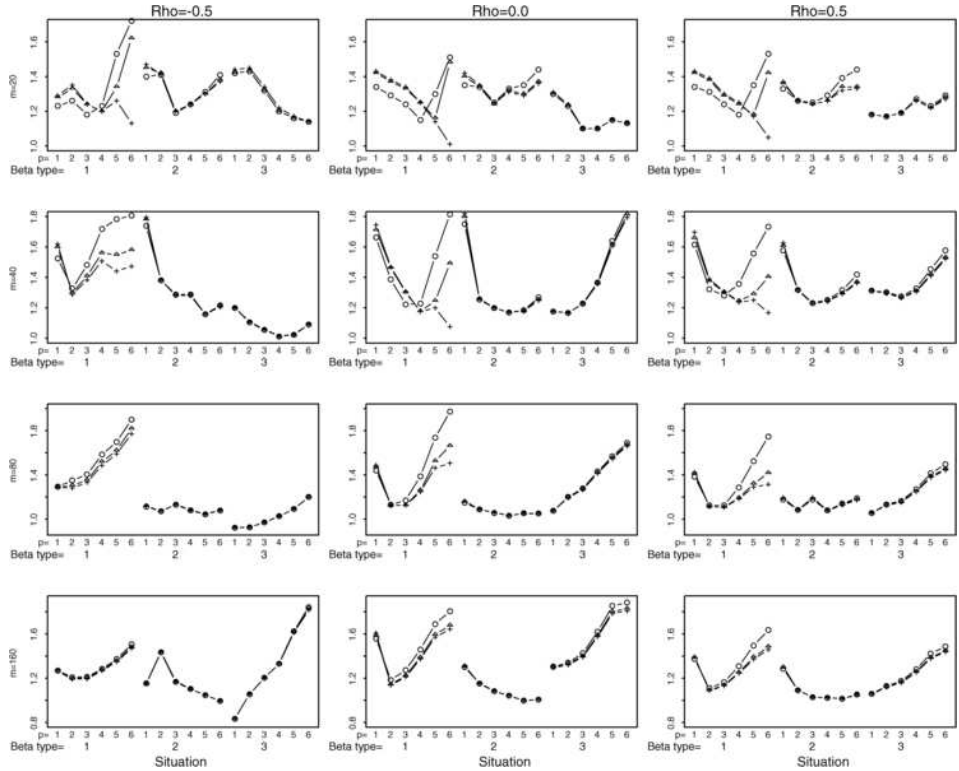

Fig. 3. *The MSPE values relative to the random oracle of FDR procedures: BH—solid line; TSFDR—dotted line; MSFDR—dashed line at 5% (○), 10% (△) and 25% (+) FDR level.*

TABLE 1
*The preferable values of q for the FDR procedures studied*

|  |  | **BH** | **TSFDR** | **MSFDR** |
|---|---|---|---|---|
| $m = 20$ and $40$ | $\rho = -0.5$ | 0.05 | 0.05 | 0.05 |
|  | $\rho = 0$ | 0.05 | 0.05 | 0.05 |
|  | $\rho = 0.5$ | 0.05 | 0.05 | 0.05 |
|  | any $\rho$ | 0.05 | 0.05 | 0.05 |
| $m = 80$ | $\rho = -0.5$ | 0.1 | 0.1 | 0.1 |
|  | $\rho = 0$ | 0.1 | 0.05 | 0.05 |
|  | $\rho = 0.5$ | 0.1 | 0.05 | 0.05 |
|  | any $\rho$ | 0.1 | 0.1 | 0.05 |
| $m = 160$ | $\rho = -0.5$ | 0.25 | 0.25 | 0.25/0.1 |
|  | $\rho = 0$ | 0.1 | 0.1 | 0.05 |
|  | $\rho = 0.5$ | 0.1 | 0.1 | 0.1 |
|  | any $\rho$ | 0.1 | 0.1 | 0.05 |

type = "2" $(1/i)$ and $p = \sqrt{m}$. For $\beta$-types = "2" and "3" for all $m$ the performances of three FDR procedures are very similar with sometimes a little advantage to the MS for full model. Still, the relative loss of using the adaptive procedures instead of BH in "small" models is less than the relative loss of using BH instead of adaptive procedures in rich models. Therefore by minimax criterion we recommend to use adaptive FDR procedures.

Note that for all FDR procedures the relative loss to random oracle decreases in $m$, which means that in the large problems ($m = 160$ and above) the performances of FDR procedures will be very close to those of the random oracle. The maximal loss relatively to random oracle of BH procedure at 0.05 gets up to $1.72, 1.78, 1.98$ and $1.87$, of two-stage FDR procedure at 0.05 gets up to $1.62, 1.74, 1.82$ and $1.81$, of multiple-stage FDR procedure at 0.05 gets up to $1.47, 1.72, 1.77$ and $1.79$ for $m = 20, 40, 80, 160$, respectively.

Based on the above, for the range of configurations investigated here, the multiple-stage FDR procedure (MSFDR) at 0.05 emerges as the recommended overall choice.

Comparing this MSFDR with $q = 0.05$ with other selection methods that have no extra tuning parameter, the empirical minimaxity summaries are given in Table 2. The two leading procedures are MSFDR 0.05 and the TK, each one achieving minimax performance for two problem sizes. Still, comparing just these two, the inefficiency of the MSFDR relative to TK is maximally 2.9%, while the other way around it is 12.9%, and for $m = 40$ the two are hardly distinguishable.

The maximal relative loss by DJ procedure gets up 3.05. It performs better than the penalties that depend both on $m$ and $p$ that reach 3.62 for FS and 6.61 for BM. Cp and FWD at 0.05 show poor performance, with 6.88 and 3.59. Since the performance deteriorates as $m$ increases, it indicates even worse performance for large problems.

TABLE 2
*The maximal relative loss (MSPE of method divided by MSPE of the random oracle). Bold figures indicate the minimax relative loss (or to within one simulation standard error). Simulation standard errors are given in parentheses*

| Procedure | $m = 20$ | $m = 40$ | $m = 80$ | $m = 160$ |
|---|---|---|---|---|
| FWD | 2.80 (0.068) | 2.87 (0.061) | 2.84 (0.043) | 3.59 (0.098) |
| Cp | 4.77 (0.096) | 4.87 (0.096) | 4.88 (0.069) | 6.88 (0.193) |
| DJ | 2.34 (0.022) | 2.78 (0.021) | 3.05 (0.016) | 2.58 (0.010) |
| BM | 4.47 (0.064) | 5.15 (0.106) | 6.61 (0.137) | 5.09 (0.110) |
| FS | 3.07 (0.097) | 3.62 (0.090) | 3.35 (0.059) | 3.35 (0.068) |
| TK | 1.66 (0.028) | **1.71** (0.015) | **1.72** (0.010) | 1.99 (0.010) |
| MSFDR 0.05 | **1.47** (0.031) | **1.72** (0.012) | 1.77 (0.010) | **1.79** (0.008) |

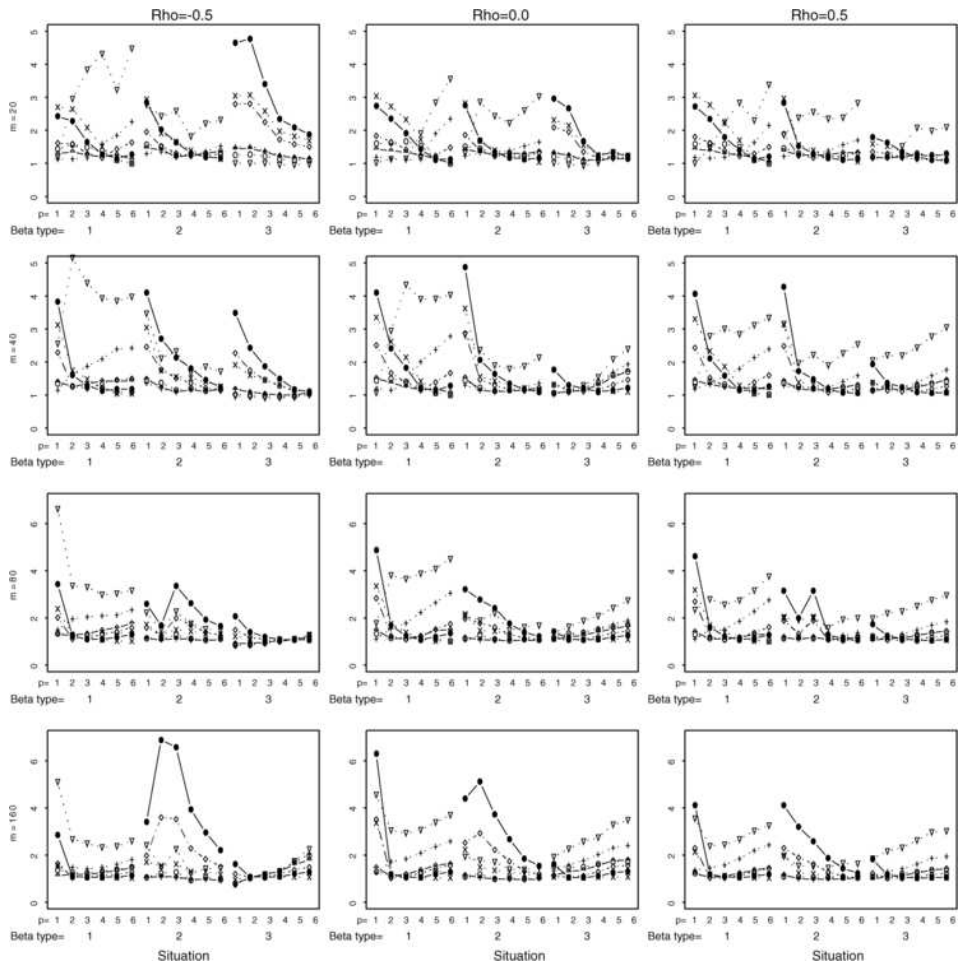

Fig. 4. *MSPE values relative to the random oracle: MSFDR ($\triangle$), DJ (+), FS ($\times$), BM ($\triangledown$), TK ($\circ$), FWD 5% ($\diamond$), Cp ($\bullet$).*

So far we have done the comparisons only utilizing the information about $m$. If further information is available on the structure of the dependency among the explanatory variables we may wish to look into the results on the finer resolution.

Figure 4 (as well as Tables 3–6 in Supplementary Materials) summarizes for each $m$ and $\rho$ separately the mean MSPE values relative to those of the random oracle over the subset of configurations of interest as discussed in Section 6. The results are presented for the MSFDR at 0.05 as the single FDR candidate and other selection procedures discussed in Section 2.2.

Again, MSFDR and the TK are the minimax procedures, about half "wins" for each, with MSFDR leading for $m = 20$ and $m = 160$, the TK

TABLE 3
*Mean relative MSPE values for the k least-favorable configurations, k = 2, 3, ALL.*
*The case for k = 1 is in Table 2*

| k | $m = 20$ | | | $m = 40$ | | | $m = 80$ | | | $m = 160$ | | |
|---|---|---|---|---|---|---|---|---|---|---|---|---|
| **Procedure** | **2** | **3** | **ALL** | **2** | **3** | **ALL** | **2** | **3** | **ALL** | **2** | **3** | **ALL** |
| FWD | 2.80 | 2.62 | 1.48 | 2.69 | 2.62 | 1.46 | 2.77 | 2.52 | 1.37 | 3.55 | 3.54 | 1.54 |
| Cp | 4.71 | 4.28 | 1.82 | 4.58 | 4.42 | 1.80 | 4.75 | 4.32 | 1.69 | 6.73 | 6.58 | 2.07 |
| DJ | 2.30 | 2.25 | 1.36 | 2.60 | 2.53 | 1.44 | 2.91 | 2.82 | 1.47 | 2.51 | 2.47 | 1.46 |
| BM | 4.39 | 4.20 | 2.03 | 4.77 | 4.63 | 2.41 | 5.55 | 5.06 | 2.35 | 4.81 | 4.44 | 2.30 |
| FS | 3.07 | 3.06 | 1.77 | 3.49 | 3.42 | 1.57 | 3.26 | 2.97 | 1.33 | 2.79 | 2.57 | 1.24 |
| TK | 1.64 | 1.63 | 1.30 | 1.66 | 1.60 | 1.23 | 1.67 | 1.61 | 1.19 | 1.86 | 1.81 | 1.23 |
| MSFDR 0.05 | 1.46 | 1.45 | 1.27 | 1.63 | 1.60 | 1.25 | 1.72 | 1.67 | 1.21 | 1.79 | 1.78 | 1.21 |

for the other two, with very small differences for $m = 40$. If we further allow $q$ to depend on $m$ and $\rho$, then when the TK has the advantage the two have similar performance and when the MSFDR takes the lead the TK is even further behind. The least-favorable situations for both were usually the extreme models either the smallest or the largest. The structure of the coefficients was either a constant or decay as the square root.

The performance of the DJ procedure is similar for all values of $\rho$. Large true models strain its performance, as can be expected. It performs better for sparse models although it is never the best, as predicted by the asymptotic theory.

The performance of BM procedure was the worst among the procedures with nonconstant penalty, although in some configurations this procedure was the best among the tested ones. Its maximal relative loss occurs for the configurations where coefficients decay as a square root, mostly for extreme size of true model: too sparse or too rich.

GF procedure performs poorly in spite of its adaptive penalty, with a more than twentyfold increase in MSPE (and therefore is not presented in Figure 4), because it is not high enough for the first variables that enter the model. In addition, because the penalty starts to decrease as the model includes more than $m/2$ variables, if $m/2$ explanatory variables enter the model all variables will.

To make a more comprehensive analysis of the simulation results and, in particular, to avoid the dependence of the conclusions on a single least-favorable case, we calculate the mean relative MSPE values for the $k$ least-favorable cases, $k = 2, 3$ finally for all tested configurations. The maximal means of each tested methods are presented in Table 3.

As the relative MSPE is averaged over more configurations the maximal risk and the gaps between the methods decrease. Still, there are two clear leaders: the TK and the multiple-stage procedure. Their relationship reflects the one discussed using the single least-favorable analysis.

**8. Discussion.** Recall that the minimax properties of the BH procedure were proved in ABDJ for large $m$ (asymptotically), for orthogonal variables and for sparse signals. From the simulation study it seems that these theoretical properties carry over to finite cases with and without the orthogonality of the explanatory variables for sparse signals, a property shared by all other FDR procedures. It is almost as good as the performance of the random oracle, which means that given a random sequence of the nested models the FDR procedures do the "achievable best." Interestingly, in the scope of configurations studied in our simulation, using FDR level of 0.05 is global best.

When we enlarge the scope of configurations to nonsparse as well as sparse the multiple-stage FDR procedure at level 0.05 turns out to be the best among FDR procedures. Comparing this single FDR controlling procedure with other penalized methods, there are two procedures that perform well across all tested situations: TK and the multiple-stage FDR procedure at level 0.05, with an advantage to the multiple-stage FDR at large problems.

Why is the multiple-stage FDR procedure so successful? We think that it can be attributed to its good frequentist behavior for sparse signals, where it behaves like the BH procedure that adapts to the level of sparsity in the right way. At the other extreme, where $p/m$ may be close to 1, it becomes close to an empirical Bayes rule, where $p/m$ is "estimated" by approximately $(1 - q)$ times the proportion of coefficients currently in the model, and then the Bayes rule is applied. This is achieved while keeping an inherently smooth transition from one situation to the other. This can be compared to the approach of Johnstone and Silverman (2005), who combine the two approaches for model selection in a different way.

In order to understand the similarity and differences in the results for the TK and MSFDR procedures, Figure 5 compares the penalty per coefficient of the two for $m = 160$. The penalty of the MSFDR procedure is lower for the first variables that enter the model than the penalty of the TK (if $m > 5$), and this gap grows up as $m$ increases. In fact, the ratio between two penalties at the first step gets close to 1.8 for $m = 10,000$. The cutoff point is at about $m/5.5$. Above the intersection point the gap between the TK and MSFDR penalties decreases as $m$ increases. Since both procedures are used in a step-forward manner, the performance of TK procedure is expected to be worse than that of the multiple-stage procedure when the model does not contain variables with extremely strong signals.

To strengthen the previous conclusion we extend our simulation study. In the new simulation we compare just the two leading methods: MS and TK procedures for 500 explanatory variables with $p$ nonzero coefficients, where $p = \sqrt{m} \approx 25$, $p = m/2 = 250$ and $p = m = 500$. The coefficients were chosen constant, so that theoretical $R^2 = 0.75$. (It was mostly a least-favorable case for both procedures.) The number of observations is $n = 1000$ (as before $n = 2m$). At $p = 250$ the performance was the same to within simulation standard error. At $p = 500$ the TK had some advantage, and at $p \approx \sqrt{m}$ the MS had a large advantage, so that the maximal MSPE of MS relative to TK turned out to be 0.82.

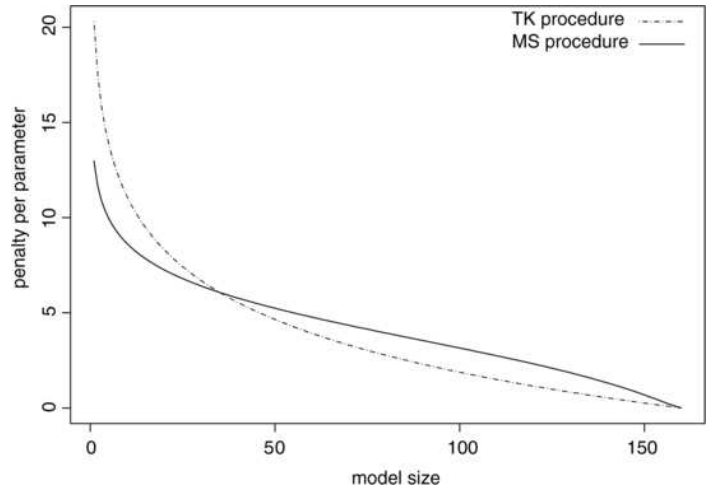

Fig. 5. *Comparing the multiple-stage FDR and the Tibshirani–Knight penalties at the studied problem sizes for $m = 160$. The solid line for MSFDR at 0.05 and the dotted line for TK procedure.*

As noted before, allowing dependency of $q$ on $m$ in the MSFDR is not a limitation since $m$ is always known and does not depend on the data structure. The simulation study in ABDJ for the orthogonal case studied the BH performance for $m = 1024$ and 65,536, and the recommended value was about $q = 0.4$. It indicates that the optimal choice of $q$ may increases in $m$ (but note that this need not be the case for the minimax relative to the oracle loss). The dependency of $q$ on $m$, and possibly on other factors known to the modeler, is an interesting research problem that will benefit from further theoretical and empirical study.

A problem that has not been addressed in this article is the challenging case where the number of potential variables $m$ is larger than the number of observations $n$. Our initial simulations show good performance for the MS method in this case as well, assuming that $\sigma^2$ is known. While if $m < n$ the standard deviation can be estimated from the full model, if $m \geq n$ the estimation of the standard deviation is a real challenge. We therefore leave this problem for future work.

## SUPPLEMENTARY MATERIAL

**Supplementary Materials** (DOI: [10.1214/08-AOAS194SUPP](); .pdf). We present the configurations used to date in comparative studies of model selection methods, and give in detail the set of configurations used in the current comparative analysis of the penalty based model selection methods. We then give the detailed results of the analysis, the summary of which is presented in the paper.

DEPARTMENT OF STATISTICS
& OPERATIONS RESEARCH
TEL AVIV UNIVERSITY
TEL AVIV 69978
ISRAEL
E-MAIL: ybenja@post.tau.ac.il
gyulia@post.tau.ac.il